\documentclass[aps,prb,twocolumn,showpacs,superscriptaddress]{revtex4}

\usepackage{graphicx}
\usepackage{bm}
\usepackage{amsmath}
\usepackage{dcolumn}%

\newcommand{\SL}{YBCO/LCMO SL}
\newcommand{\OBa}{O$_{\text{Ba}}$}
\newcommand{\OLa}{O$_{\text{La}}$}
\newcommand{\OCu}{O$_{\text{Cu}}$}
\newcommand{\OMn}{O$_{\text{Mn}}$}
\newcommand{\MMn}{\ensuremath{M_{\text{Mn}}}}
\newcommand{\MCu}{\ensuremath{M_{\text{Cu}}}}
\newcommand{\mb}{\ensuremath{\mu_{\text{B}}}}
\newcommand{\ef}{\ensuremath{\varepsilon_{F}}}

\newcommand{\eg}{\ensuremath{e_{g}}}
\newcommand{\tg}{\ensuremath{t_{2g}}}
\newcommand{\dxy}{\ensuremath{d_{xy}}} 
\newcommand{\dxz}{\ensuremath{d_{zx}}}
\newcommand{\dyz}{\ensuremath{d_{yz}}}
\newcommand{\dxyz}{\ensuremath{d_{xz,yz}}}
\newcommand{\dzz}{\ensuremath{d_{3z^2-1}}}
\newcommand{\dxxyy}{\ensuremath{d_{x^2-y^2}}}
\newcommand{\pxy}{\ensuremath{p_{x,y}}}
\newcommand{\pz}{\ensuremath{p_{z}}}

\newcommand{\duxy}{\ensuremath{d_{\uparrow xy}}}
\newcommand{\duxz}{\ensuremath{d_{\uparrow zx}}}
\newcommand{\duyz}{\ensuremath{d_{\uparrow yz}}}
\newcommand{\duzz}{\ensuremath{d_{\uparrow 3z^2-1}}}
\newcommand{\duxxyy}{\ensuremath{d_{\uparrow x^2-y^2}}}
\newcommand{\ddxy}{\ensuremath{d_{\downarrow xy}}}
\newcommand{\ddxz}{\ensuremath{d_{\downarrow zx}}}
\newcommand{\ddyz}{\ensuremath{d_{\downarrow yz}}}
\newcommand{\ddzz}{\ensuremath{d_{\downarrow 3z^2-1}}}
\newcommand{\ddxxyy}{\ensuremath{d_{\downarrow x^2-y^2}}}

\newcolumntype{/}{D{/}{/}{2,2}}  
\newcolumntype{.}{D{.}{.}{-1}}  

\begin{document}

\title{Electronic structure and x-ray magnetic circular dichroism of
  YBa$_2$Cu$_3$O$_7$/La$_{1-x}$Ca$_x$MnO$_3$ superlattices from
  first-principles calculations}

\author{Xiaoping Yang}
\affiliation{Max-Planck-Institut f\"ur Festk\"orperforschung,
  Heisenbergstra{\ss}e 1, D-70569 Stuttgart, Germany}

\author{A.~N. Yaresko}
\affiliation{Max-Planck-Institut f\"ur Festk\"orperforschung,
  Heisenbergstra{\ss}e 1, D-70569 Stuttgart, Germany}

\author{V.~N. Antonov}
\affiliation{Max-Planck-Institut f\"ur Festk\"orperforschung,
  Heisenbergstra{\ss}e 1, D-70569 Stuttgart, Germany}
\affiliation{Institute of Metal Physics, 36 Vernadsky Street, 03142
Kiev, Ukraine}

\author{O.~K. Andersen}
\affiliation{Max-Planck-Institut f\"ur Festk\"orperforschung,
  Heisenbergstra{\ss}e 1, D-70569 Stuttgart, Germany}

\date{\today}

\begin{abstract}

  The origin of x-ray magnetic circular dichroism (XMCD) at the Cu $L_{2,3}$
  edge in YBa$_2$Cu$_3$O$_7$/La$_{1-x}$Ca$_x$MnO$_3$ superlattices is revealed
  by performing first-principle electronic structure calculation using
  fully-relativistic spin-polarized linear muffin-tin orbital and projected
  augmented plane wave methods. We show that the XMCD spectra at the Cu
  L$_{2,3}$ edges are proportional to the difference of the densities of Cu
  \duzz\ and \ddzz\ states. Although the Cu \dzz\ states lie well below the
  Fermi level, a small number of \duzz\ holes is created by the Cu \dzz--\OBa\
  $p_z$--Mn \dzz\ hybridization across the interface. Even this tiny number of
  holes is sufficient to produce appreciable Cu L$_{2,3}$ XMCD. The robustness
  of this conclusion is verified by studying the influence of doping, atomic
  relaxation, correlation effects, and antiferromagnetic order in a CuO$_2$
  plane on the XMCD spectra.

\end{abstract}


\maketitle

\section{\label{sec:introd}Introduction}

Artificially grown heterostructures and superlattices of alternating
superconducting (SC) and ferromagnetic (FM) materials have become an important
tool for exploring the interplay between superconductivity and
ferromagnetism. Various interaction phenomena that may occur, like proximity
effects, coupling phenomena, charge transfer etc., are directly influenced by
the structure and especially by the physical and/or chemical disorder of the
interfaces that separate them. The combination of ferromagnetic and
superconducting materials could end up in devices like a spin-controlled
transistor with a high gain current and short switching times.

In the last years much attention has been payed to oxide based FM/SC
superlattices (SL), combining a high-$T_c$ superconductor
YBa$_2$Cu$_3$O$_7$ (YBCO) with a La$_{2/3}$Ca$_{1/3}$MnO$_3$ (LCMO) manganite
which exhibits colossal
magnetoresistance. \cite{SVP+02,VLP+03,SAP+03,HHC+04,HGS+04,SCN+05,ADS+07,PGS+07,FCH+07,CFS+06,CFH+07}
Both YBCO and LCMO have oxide perovskite structure with very similar in-plane
lattice parameters, which allows the growth of superlattices with sharp
interfaces, thus strongly reducing extrinsic "structural" effects which
otherwise could obscure the FM/SC interplay. Using scanning transmission
electron microscopy and high spatial resolution electron energy loss
spectroscopy, Varela {\it et al.} showed in Ref.~\onlinecite{VLP+03} that
individual layers of the YBa$_2$Cu$_3$O$_7$/La$_{1-x}$Ca$_x$MnO$_3$
superlattices are flat over long lateral distances. The interfaces are
coherent, free of defects, exhibiting no roughness, and are located at the BaO
plane of the superconductor. Concerning chemical disorder, EELS measurements
show the absence of measurable chemical interdiffusion within experimental
error bars.

The work on the perovskite oxide FM/SC superlattices is motivated by the
appealing properties of the cuprate high-$T_c$ superconductors (HTSC) whose
high SC critical temperatures make them potentially useful for technological
applications. Further, since HTSC are believed to be susceptible to a variety
of competing instabilities, there is a high potential for SC/FM quantum states
in multilayer structures. It has been suggested by Si in
Ref.~\onlinecite{Si97} that electron injection from hole-doped rare earth
manganites into high-temperature superconductors can yield information on
spin-charge separation in high-$T_c$ superconductors. The spin-polarization,
in addition to the ferromagnetic ordering nature, is one particular property
of hole-doped rare earth $R_{1-x}A_x$MnO$_3$ manganite, where $R$ is a
trivalent rare-earth ion and $A$ is a divalent alkaline-earth ion. In the case
of La$_{2/3}$Ca$_{1/3}$MnO$_3$ a nearly full spin polarization of the
transport electrons can be found. \cite{SBO+98}

Magnetization measurements of epitaxially grown bilayers of LCMO and YBCO show
a coexistence of ferromagnetism and superconductivity in these bilayer samples
at low temperatures. \cite{SAC+04} Both magnetization measurements which are
used to determine $T_c$ and magneto-optical measurements for the $j_c$
evaluation of the superconducting film show up a critical thickness for the
YBCO film of about $d$=30 nm, below which the superconductivity is strongly
affected by the magnetic film. Sefrioui {\it et al.}  in
Ref.~\onlinecite{SAP+03} studied the interplay between magnetism and
superconductivity in high-quality YBCO/LCMO superlattices. They found
evidences for the YBCO superconductivity depression in the presence of the
LCMO layers. Due to its short coherence length, superconductivity survives in
the YBCO down to a much smaller thickness in the presence of the magnetic
layer than in low $T_c$ superconductors. For a fixed thickness of the
superconducting layer, superconductivity is depressed over a thickness
interval of the magnetic layer in the 100 nm range.

Holden {\it et al.} in Ref.~\onlinecite{HHC+04} reported ellipsometry
measurements of the far-infrared dielectric properties of SL composed of thin
layers of YBCO and LCMO. The optical data provide clear evidence that the
free-carrier response is strongly suppressed in these SL as compared to the
one in the pure YBCO and LCMO films. The suppression occurs in the normal as
well as in the SC state and involves a quite large length scale.

This research is in its early stage, and relatively little is known about the
nature of magnetism at the interface, the spatial distribution of the
magnetization throughout the layers, and the interplay of FM and SC order
parameters in general.

X-ray magnetic circular dichroism (XMCD) spectra at Mn and Cu $L_{2,3}$ edges
in YBCO/LCMO SL were measured by Chakhalian {\it et al.} in
Ref.~\onlinecite{CFS+06}. It was shown that magnetic dichroism is clearly
present at both Mn and Cu edges, although the Cu XMCD signal is small compared
to that of Mn (27\%) and does not exceed 1.4\%.  The measurements suggest the
presence of an uncompensated induced magnetic moment in a YBCO layer close to
the YBCO/LCMO interface. As manganite layers undergo a ferromagnetic
transition at around 180 K, the large dichroism at the Mn $L_{2,3}$ edge is
expected. However, the presence of the net ferromagnetic magnetization on Cu is
quite surprising. Authors investigated also the temperature dependence of the
dichroic signals at both the Mn and Cu $L_{2,3}$ edges. At temperatures above
170 K, the dichroism on Mn disappears, consistent with the
ferromagnet-paramagnet transition. Similar behavior was also found for Cu,
except that already above 150 K its XMCD signal decreases below the detection
limit. From the similar temperature dependences of the XMCD spectra authors
concluded that the magnetic moment on Cu is induced by strong interactions
between Cu spins and the ferromagnetic moment of Mn across the interface.

Later, x-ray absorption spectra (XAS) and x-ray linear dichroism (XLD) at Mn
and Cu $L_{2,3}$ edges were recorded in the total electron yield (TEY) and
fluorescence-yield (FY) modes from the same sample of YBCO/LCMO
SL. \cite{CFH+07} In contrast to strong Cu $L_{2,3}$ XLD detected in the bulk
sensitive FY mode, only a very weak XLD signal was measured in the TEY mode,
which is supposed to probe interfacial Cu ions. This led authors to the
conclusion that the electronic states of Cu ions at the YBCO/LCMO interface
are strongly reconstructed, with the number of holes in the Cu \dzz\ states
being close to the number of Cu \dxxyy\ holes. A mechanism for such orbital
reconstruction was proposed.

In the present study, we focus our attention on the electronic band structure
of YBa$_2$Cu$_3$O$_7$/La$_{1-x}$Ca$_x$MnO$_3$ superlattices and their XMCD and
XAS spectra. We show that the magnetic circular dichroism at the Cu L$_{2,3}$
edges appears due to the hybridization between completely filled Cu \dzz\
states and unoccupied Mn \dzz\ states split by the on-site exchange
interaction. The number of Cu \dzz\ holes created by the hybridization is much
smaller than the number of holes in the Cu \dxxyy\ states. Nevertheless, even
this tiny number of \dzz\ holes is sufficient to produce appreciable Cu
L$_{2,3}$ XMCD. The robustness of the results is verified by studying the
influence of doping, atomic relaxation and correlation effects on the
electronic structure and XMCD spectra of YBCO/LCMO SL.

The paper is organized as follows. The details of the calculations are
described in Sec.~\ref{sec:details}. A structural model of YBCO/LCMO
superlattices considered in this study is presented in Sec.~\ref{sec:str}.
Section \ref{sec:lsda} is devoted to the electronic structure and XMCD spectra
of the YBCO/LCMO SL calculated using the local spin-density approximation
(LSDA) and the LSDA+$U$ method.  Finally, the results are summarized in
Sec.~\ref{sec:summ}.

\section{\label{sec:details}Computational details}

The electronic structure and XMCD spectra of \SL\ were investigated by means
of \textit{ab initio} calculations based on the spin-polarized density
functional theory. The energy band structure, densities of states (DOS), and
magnetic moments were calculated within the local spin-density approximation
(LSDA) using the projected augmented plane wave (PAW) method
\cite{Bl94,PBE97,KJ99} and the linear muffin-tin orbital (LMTO) method.
\cite{And75} The cutoff energy of the plane-wave expansion was 500 eV and the
spacing between $k$ points was 0.03 {\AA}$^{-1}$.

XMCD calculations were performed using the spin-polarized fully relativistic
LMTO method (see Refs.~\onlinecite{APSY95,book:YPAH06} for details). To
investigate the effect of electronic correlations on the XMCD spectra the
rotationally invariant LSDA+$U$ method \cite{LAZ95,YAF03} with the
double-counting term approximated by the atomic limit\cite{CS94} was used. The
screened on-site Coulomb repulsion $U$ was considered as an adjustable
parameter. We used $U$=3 to 5 eV for the Mn $d$ states and $U$=7.5 eV for the
Cu $d$ states. For the exchange integral $J$ the value of 1.0 eV estimated
from constrained LSDA calculations was used.

The finite lifetime of a core hole was accounted for by folding the XMCD
spectra with a Lorentzian. The widths of the $L_2$, and $L_3$ core level
spectra, with $\Gamma_{L_2}$=0.97 eV, $\Gamma_{L_3}$=0.36 eV for the Mn and
$\Gamma_{L_2}$=1.04 eV, $\Gamma_{L_3}$=0.61 eV for the Cu were taken from
Ref.~\onlinecite{CaPa01}. Finite instrumental resolution was accounted for by
a Gaussian of 0.5 eV.

\section{\label{sec:str}Structural model}

\begin{figure}[tbp!]
\begin{center}
\includegraphics[width=0.30\textwidth]{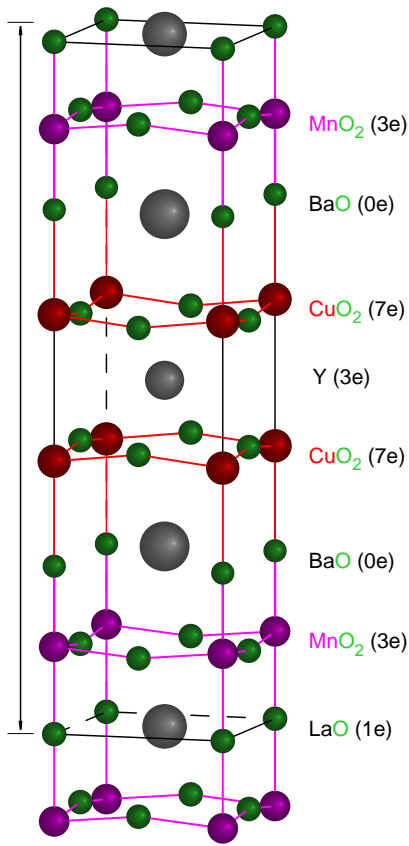}
\end{center}
\caption{\label{fig:struc}(Color online) Schematic representation of the
  crystal structure of the $1\times$YBaCuO/$1\times$LMO superlattice. The
  tetragonal unit cell used in the calculations is marked by the
  double-arrowed line. The number of electrons provided by each plane to Cu
  and Mn $d$ shells is given in parenthesis. }
\end{figure}

The crystal structure of \SL\ was approximated by a tetragonal ($P4/mmm$) unit
cell consisting of one YBCO and up to 3 LMO unit cells stacked along the $c$
direction. The interface is formed by a BaO plane from the YBCO side and a
MnO$_2$ plane of LMO, which replaces a CuO plane of YBCO. The unit cell
consisting of one YBCO and one LMO layer is shown in Fig.~\ref{fig:struc}. For
the in-plane lattice constants $a$=$b$=3.94 \AA\ the experimental value for
pseudo-cubic LMO is used. Thus, the orthorhombic distortion of bulk YBCO is
neglected and the YBCO layer is slightly expanded in the $ab$ plane.  The $c$
constant of 15.61 \AA\ is taken to be equal to the sum of the lattice
constants of bulk YBCO ($c_{\mathrm{YBCO}}$=11.67 \AA) and pseudo-cubic LMO
($c_{\mathrm{LMO}}$=3.94 \AA).

Interfacial ions are subject to forces different from those in the bulk and,
thus, can change their positions. The mere transfer of charge at an interface
can lead to "electronic reconstruction". \cite{OkMi04} Since the exact atomic
structure at the YBCO/LCMO interface is not known, atomic positions were
optimized using the PAW method. During the optimization the $a$ and $c$
lattice constants were fixed to the above mentioned values. As can be
expected, the most sensitive to the optimization are the positions of the
atoms which are close to the YBCO/LCMO interface. An oxygen ion denoted as
\OBa, which sits in the BaO plane between the nearest Cu and Mn ions, moves in
the course of the optimization towards Mn. As a result, the Cu--\OBa\ distance
increases from 2.29 \AA\ in bulk YBCO to 2.37 \AA\ in the superlattice and,
thus, becomes even larger as compared to the separation of 1.98 \AA\ between
Cu and \OCu\ ions within the CuO$_2$ plane. The Mn--\OBa\ distance decreases
from 1.97 \AA\ to 1.82 \AA. The optimized separation between Cu and Mn ions is
slightly (4.18 \AA\ vs.\ 4.15 \AA) larger than between Cu ions residing in
CuO$_2$ and CuO planes in bulk YBCO.  The structure optimization leads also to
weak buckling of MnO$_2$ planes. Nevertheless, the Mn--\OMn\ distance (1.98
\AA) remains practically the same as the Mn--\OLa\ one (1.97 \AA). As the LaO
plane is a mirror plane, the positions of \OLa\ ions do not change during the
optimization. Thus, the interfacial Mn ion is surrounded by a slightly
distorted O$_6$ octahedron formed by \OBa, \OLa, and 4$\times$\OMn\ ions.

Atomic positions were also optimized for $1\times$YBCO/$2\times$LMO SL with
$c$=19.55 \AA\ ($c_{\mathrm{YBCO}}+2c_{\mathrm{LMO}}$) containing 2 LMO unit
cells. In this case, the Cu--\OBa\ and Mn--\OBa\ distances of 2.48 \AA\ and
1.80 \AA\ were obtained, which gives the separation of 4.28 \AA\ between Cu
and Mn ions at the interface. As the $c$ lattice constant was fixed, the
expansion at the interface occurs at the expense of some decrease of the
distances between CuO$_2$ planes and between the ``bulk'' and ``interfacial''
MnO$_2$ planes.

Under the assumption that O $p$ states are completely filled and $s$ and $d$
states of Ba, Y, and La are empty, Mn and Cu $d$ states in the SL shown in
Fig.~\ref{fig:struc} are filled with 24 electrons. Taking into account that
the unit cell contains 2 Cu and 2 Mn ions and assuming that some charge
$\delta$ may be transferred across the interface, the occupations of the $d$
shells are Cu $d^{9\pm\delta}$ and Mn $d^{3\mp\delta}$. Hence, the occupation
of Mn $d$ states in this model SL corresponds to La$_{1-x}$Ca$_x$MnO$_3$ with
$x\sim 1$. With each LMO layer, consisting of one MnO$_2$ and one LaO plane,
added to a model unit cell, the number of electrons which fill the 
$d$ states of transition metal ions 
increases by 4, while the number of Mn ions only by 1. Consequently, the
average occupation of Mn $d$ states increases with the increase of the LMO
layer thickness. This should be taken into account when comparing the results
calculated for different model YBCO/LMO SL.

\section{\label{sec:lsda}Results and discussion}

\subsection{\label{sec:xmcd}XMCD spectra}

The x-ray magnetic circular dichroism in the YBCO/LCMO SL has been measured at
the Mn and Cu $L_{2,3}$ edges by Chakhalian \textit{et al.}
(Ref.~\onlinecite{CFS+06}). The measurements were carried out at 30 K in an
applied field of 500 Oe on a superlattice of alternating 100 \AA-thick,
$c$-axis-oriented YBa$_2$Cu$_3$O$_7$ and La$_{2/3}$Ca$_{1/3}$MnO$_3$ layers. A
circularly polarized x-ray beam was shined at a small angle to the film
surface, so that the x-ray propagation direction was almost parallel to the
LCMO magnetization oriented in the $ab$ plane.

\begin{figure}[tbp!]
\begin{center}
\includegraphics[width=0.45\textwidth]{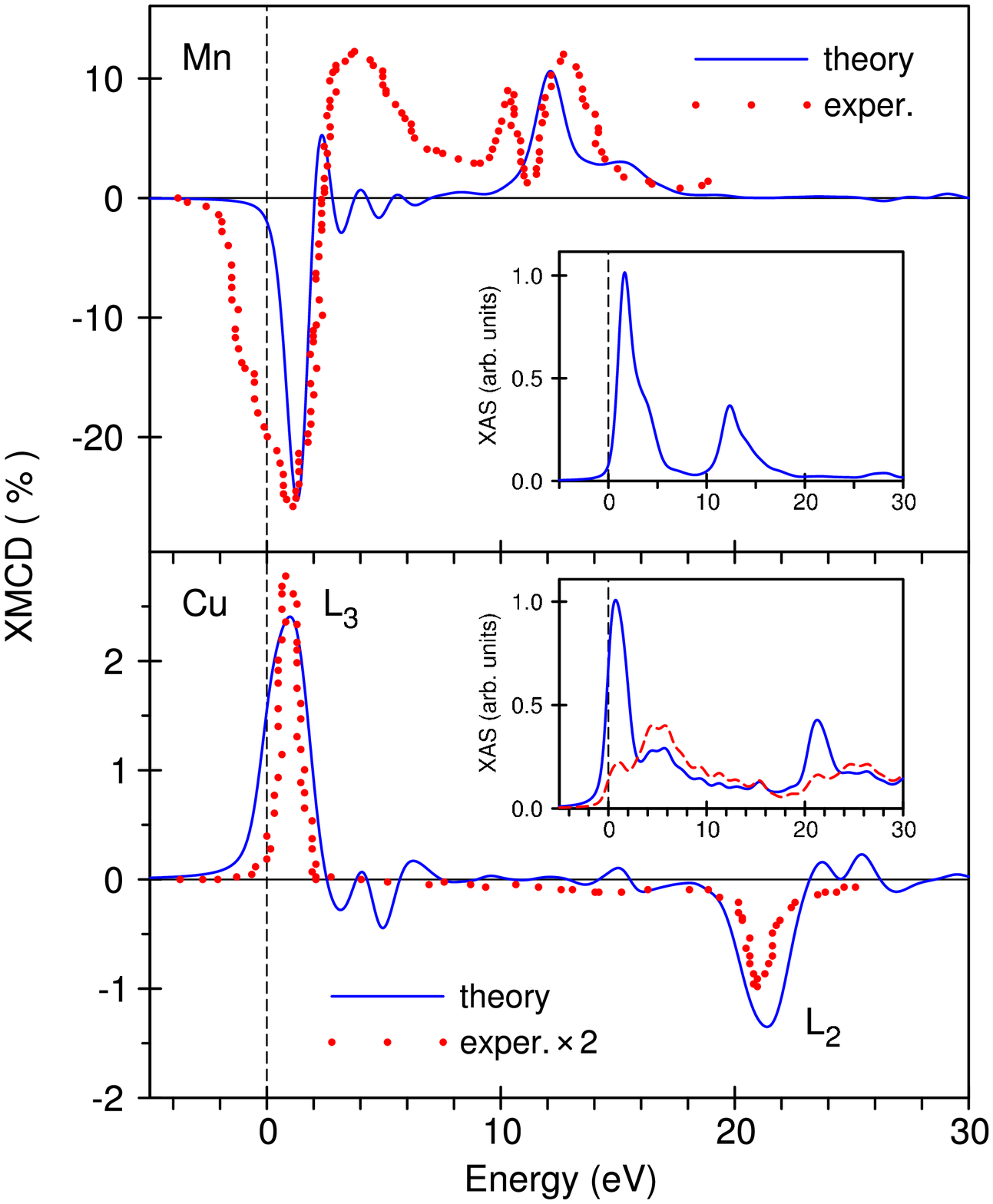}
\end{center}
\caption{\label{fig:XMCD} (Color online) The theoretical and experimental
  (Ref.~\protect\onlinecite{CFS+06}) XMCD spectra of the YBCO/LCMO
  superlattice at the Mn and Cu $L_{2,3}$ edges. The experimental Cu $L_{2,3}$
  XMCD spectrum is multiplied by a factor of 2. The insets show theoretical
  XAS spectra averaged over two circular polarizations (solid line). The Cu
  $L_{2,3}$ XAS calculated with the x-ray polarization along the $c$ axis is
  plotted by dashed line. All the spectra are plotted relative to the
  corresponding $L_3$ absorption edge marked by vertical dashed lines.}
\end{figure}

Figure \ref{fig:XMCD} shows XMCD spectra at the Mn and Cu $L_{2,3}$ edges
calculated within LSDA for the $1\times$YBCO/$1\times$LMO SL together with the
experimental spectra from Ref.~\onlinecite{CFS+06}. The corresponding
normalized theoretical x-ray absorption spectra averaged over two circular
polarizations are plotted by solid lines in the insets in
Fig.~\ref{fig:XMCD}. The LSDA calculations reproduce very well the
experimental Cu $L_{2,3}$ XMCD spectra, although the calculated magnetic
dichroism has a small oscillatory structure at the high energy part of the Cu
$L_{2,3}$ spectra which is not detected by the experiment probably due to a
weak signal. XMCD peaks at the Mn and Cu $L_3$ edges are of opposite
signs, as it is expected when the magnetization induced on Cu is antiparallel
to \MMn. The theory overestimates the dichroism at the Cu $L_{2,3}$ edges
approximately by a factor of two in comparison with the experimental
observation. This can be explained by the fact that our calculations were
performed for the unit cell containing a single YBCO cell, with the Cu ions in
both CuO$_2$ planes polarized due to their proximity to magnetic LCMO layers.
In the experimental samples, on the other hand, YBCO layers, consisting of
about 3 unit cells of bulk YBa$_2$Cu$_3$O$_7$, are significantly thicker. It
is natural to expect that only Cu ions in the interfacial CuO$_2$ planes have
a non-zero component of the magnetic moment along the light propagation
direction and, consequently, contribute to the measured XMCD signal, whereas
inner CuO$_2$ planes are XMCD-inert. With the increase of the YBCO layer
thickness in the model unit cell, the XAS intensity would increase
proportionally to the number of CuO planes, but not the dichroism. Since the
calculated XMCD spectra were normalized by the intensity of the corresponding
x-ray absorption spectra, the normalized XMCD signal would tend to
decrease with the increase of the YBCO layer thickness, which would improve
the agreement between the calculations and the experiment.

Our calculations for the $1\times$YBCO/$1\times$LMO SL result in strong linear
dichroism at the Cu $L_{2,3}$ edges. The XAS spectrum calculated for the x-ray
polarization $\varepsilon$ parallel to the $c$ axis is compared to the
spectrum averaged over two circular polarizations, which in this geometry
corresponds to the averaging over $\varepsilon||b$ and $\varepsilon||c$
polarizations, in the inset to the lower panel of Fig.~\ref{fig:XMCD}. A
prominent peak at $\sim$1 eV in the averaged spectrum is strongly suppressed
when $e$ becomes parallel to $c$. The strong XLD at the Cu $L_{2,3}$ edges is
at variance with the interface-sensitive TEY measurements, \cite{CFH+07} which
showed a very weak polarization dependence of Cu $L_{2,3}$ XAS.

In the case of Mn, the experimental $L_{2,3}$ XMCD spectrum shows two
additional positive peaks at the energies around 4 eV and 10 eV which are
absent in the theoretical spectrum. The cause of such a disagreement is yet
unclear. One of possible reasons may be many-body effects, e.g., the
interaction of the partially filled Mn $d$ states with a $2p$ core hole, which
is not taken into account in the theoretical calculations. The Mn $L_{2,3}$
XMCD spectrum may change if the magnetization in interfacial MnO$_2$ planes is
reversed relative to the rest of the LCMO layer or it may be strongly
influenced by possible orbital ordering in LCMO. These questions need further
theoretical consideration.

In the following we focus on the origin of the magnetic circular dichroism at
the Cu $L_{2,3}$ edges, but first we turn our attention to the electronic band
structure of YBCO/LMO superlattices.

\subsection{\label{sec:bands}Energy band structure}

\begin{figure}[tbp!]
\begin{center}
\includegraphics[width=0.47\textwidth]{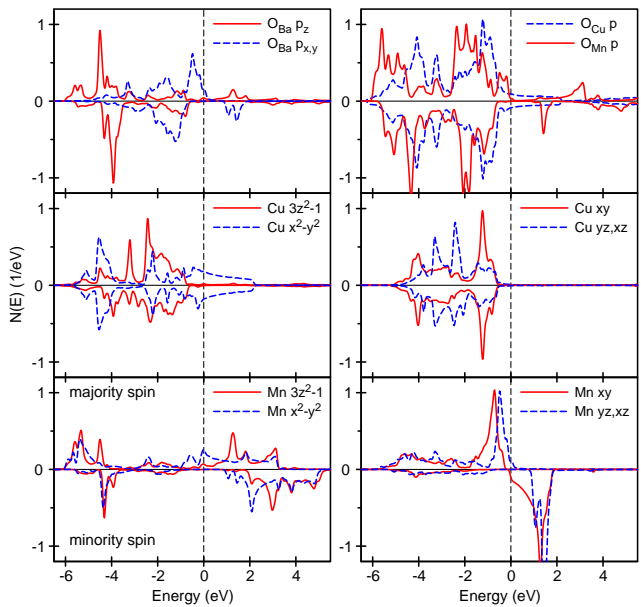}
\end{center}
\caption{\label{fig:DOS_LSDA} (Color online) Spin- and symmetry-resolved LSDA
  densities of states, $N(E)$, for $1\times$YBCO/$1\times$LMO SL. The Fermi
  energy is at zero.}
\end{figure}

\begin{figure*}[tbp!]
\begin{center}
\includegraphics[width=0.85\textwidth]{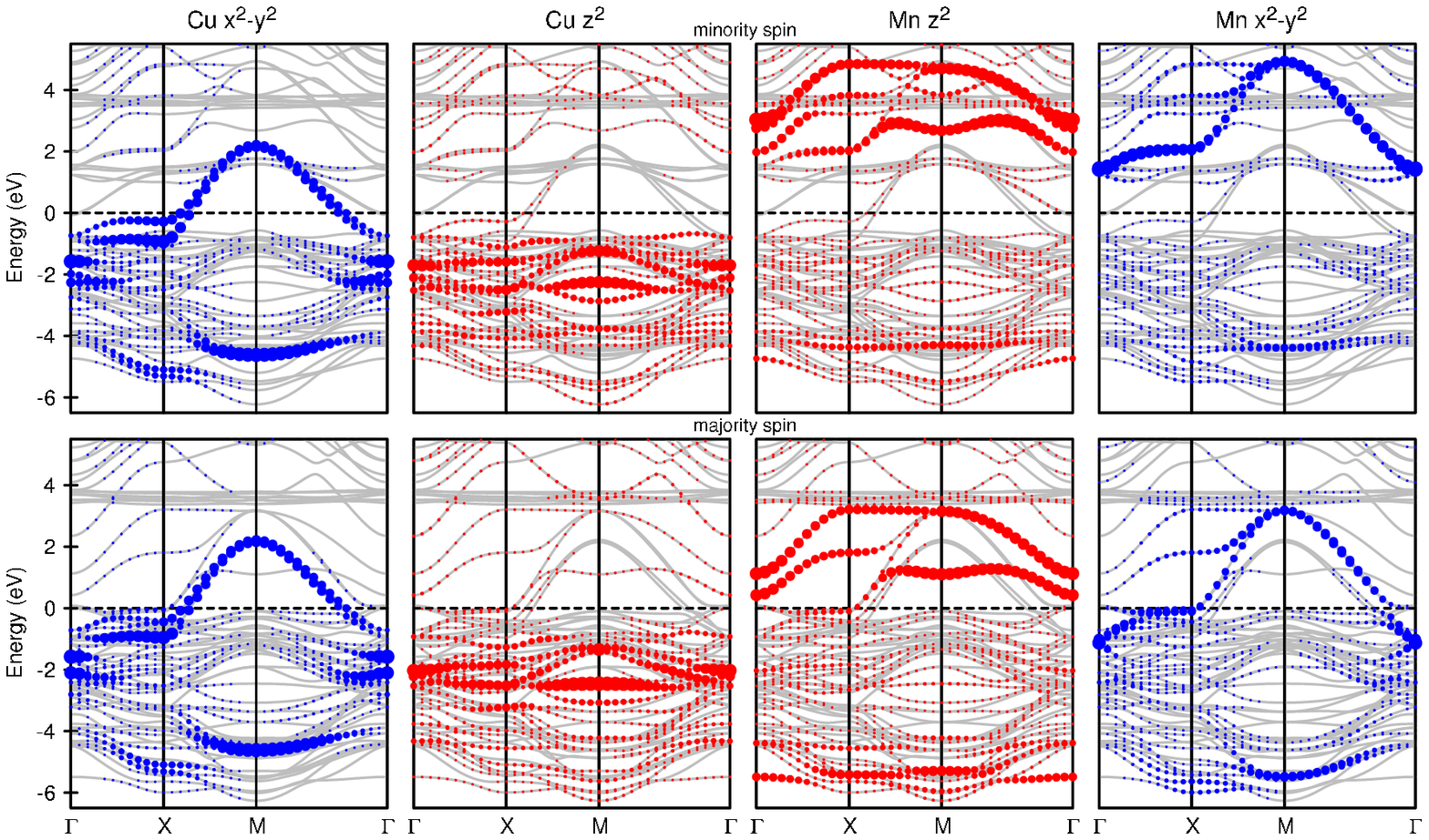}
\end{center}
\caption{(Color online) The energy band dispersion in the fat-band
  representation calculated within LSDA for the model
  $1\times$YBCO/$1\times$LMO SL shown in Fig.~\ref{fig:struc}. The size of a 
  circle is proportional to the weight of the corresponding state in a Bloch
  wave function at a given k-point.}
\label{fig:BND_FB} 
\end{figure*}

Symmetry-resolved densities of Cu $d$, Mn $d$, and O $p$ states calculated
within LSDA for the model $1\times$YBCO/$1\times$LMO SL (Fig.~\ref{fig:struc})
assuming ferromagnetic ordering of Mn magnetic moments are presented in
Fig.~\ref{fig:DOS_LSDA}. Mn $d$ states are split by the octahedral component
of the ligand field into rather narrow \dxy\ and \dxyz\ (``$t_{2g}$'') states
and much wider \dxxyy\ and \dzz\ (``$e_{g}$'') ones. The latter are strongly
hybridized with, respectively, \OMn\ $p_{x,y}$ and \OBa\ and \OLa\ $p_z$
states. The \dzz\ states, or more precisely their antibonding combinations
with O $p$ states, are shifted to higher energies due to the shortening of the
Mn--\OBa\ bond at the interface. Because of the small thickness of a LMO
layer, they are narrower than the \dxxyy\ states.

The on-site Hund's coupling splits the Mn \tg\ states into almost completely
occupied majority- and unoccupied minority-spin states. The \tg\ states
provide the dominant contribution (2.21\mb) to the magnetization
\MMn=2.75\mb\ inside a Mn sphere. The rest comes from the Mn
\dxxyy\ (0.32\mb) and \dzz (0.21\mb) states which are also split by the
on-site exchange interaction. While the minority-spin \eg\ states are empty,
the bottom of the majority-spin \dzz\ and, especially, \dxxyy\ states is
partially filled. 

Figure \ref{fig:DOS_LSDA} shows that, as it is usual for cuprates, all Cu $d$
states, except for \dxxyy\ ones, lie well below the Fermi level (\ef). The
\dxxyy\ states, interacting strongly in the $ab$ plane due to $dp\sigma$-type
bonds with \OCu\ \pxy\ states, form 3 eV--wide bands in the range from $-$1 to
2 eV, as can be seen in Fig.~\ref{fig:BND_FB} which shows the contributions of
the \dzz\ and \dxxyy\ states of Cu and Mn to the bands calculated for the
$1\times$YBCO/$1\times$LMO SL. The bands originating from Cu and Mn \dxxyy\
states have similar dispersions. The Cu \dxxyy\ bands for both spin directions
are nearly degenerate and cross \ef. The majority-spin Mn bands also cross
\ef, whereas the minority-spin ones lie 2 eV higher and are completely empty.
Mn \dzz-derived bands are split into bonding and antibonding ones by strong
interaction between the \dzz\ orbitals of Mn ions from two MnO$_2$ planes via
\OLa\ \pz\ states. The bonding majority-spin Mn \dzz\ bands are found just
above \ef.

Finally, Cu \dzz\ bands for both spin directions lie about 2 eV below \ef\ and
hybridize strongly with O $p$ bands.  Owing to the hybridization of the Cu
\dzz\ states with the spin-split Mn \dzz\ states, which is mediated by \OBa\
\pz\ orbitals, the densities of the majority- and minority-spin Cu \dzz, as
well as \OBa\ \pz, states are remarkably different (see
Fig.~\ref{fig:DOS_LSDA}), with the differences being mostly restricted to the
occupied bands. However, due to the Cu \dzz--\OBa\ \pz--Mn \dzz\
hybridization, the Cu \dzz\ states contribute also to the unoccupied Mn
\dzz-derived bands. Because of the large (2.37 \AA) Cu--\OBa\ separation and a
large energy difference between the Cu and Mn \dzz\ states, this contribution
is quite small and can hardly be seen in Fig.~\ref{fig:DOS_LSDA}. It is,
however, important for understanding the origin of Cu $L_{2,3}$ XMCD spectra
presented in Fig.~\ref{fig:XMCD}.

\begin{figure}[tbp!]
\begin{center}
\includegraphics[width=0.3\textwidth]{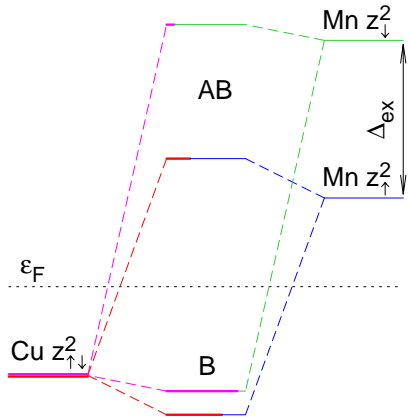}
\end{center}
\caption{(Color online) A sketch of the hybridization between the occupied Cu
  \dzz\ states and the unoccupied Mn \dzz\ states split by the on-site
  exchange interaction $\Delta_{ex}$. The contribution of the Cu \dzz\
  orbitals to bonding (B) and antibonding (AB) states is shown by thick
  lines. The negative spin-polarization of the Cu \dzz\ states appears due to
  their larger weight in the majority-spin AB states than in the minority-spin
  ones.}
\label{fig:levels} 
\end{figure}

It is well documented that LSDA fails to produce a local Cu magnetic moment in
HTSC cuprates.\cite{ZJGP+88,CS94} 
In the model $1\times$YBCO/$1\times$LMO SL, a
tiny Cu moment \MCu=$-$0.002\mb, i.e., directed oppositely to the Mn moment,
is induced by the proximity of a ferromagnetic MnO$_2$ plane. The dominant
contribution of $-$0.014\mb\ to \MCu\ is provided by the \dzz\ states. It is,
however, partially compensated by the magnetization of opposite sign
(0.012\mb) induced in the \dxyz\ chanel. The contributions of the \dxy\
(0.003\mb) and \dxxyy\ ($-$0.003\mb) states are even smaller and cancel each
other. Such a small difference in the occupations of the majority- and
minority-spin Cu $d$ states produces negligibly small exchange splitting of
the $d$ shell.
The \dzz\ states become spin-polarized as a result of their hybridization with
the spin-split unoccupied Mn \dzz\ states. The majority-spin Mn \dzz\ bands
disperse just above \ef, about 3 eV higher than the center of Cu \dzz\ states,
while the minority-spin bands are shifted by the on-site exchange splitting
$\Delta_{ex}\sim 2$ eV to higher energies (Fig.~\ref{fig:BND_FB}). Because of
this energy difference, the Cu \dzz--\OBa\ \pz--Mn \dzz\ hybridization creates
a larger number of holes in the majority- than in the minority-spin \OBa\ \pz\
and Cu \dzz\ states, as illustrated schematically in
Fig.~\ref{fig:levels}. Consequently, the sign of the corresponding
contributions to Cu and \OBa\ magnetizations is opposite to the magnetization
of Mn. The negative \OBa\ \pz\ spin-polarization is, however, over-compensated
by the positive one, induced by the hybridization of \pxy\ orbitals with Mn
\dxyz\ states: as the majority-spin \dxyz\ states are completely filled, only
minority-spin \pxy\ holes are created and the calculated total \OBa\ spin
moment of 0.13\mb\ is parallel to \MMn.

\begin{figure}[tbp!]
\begin{center}
\includegraphics[width=0.45\textwidth]{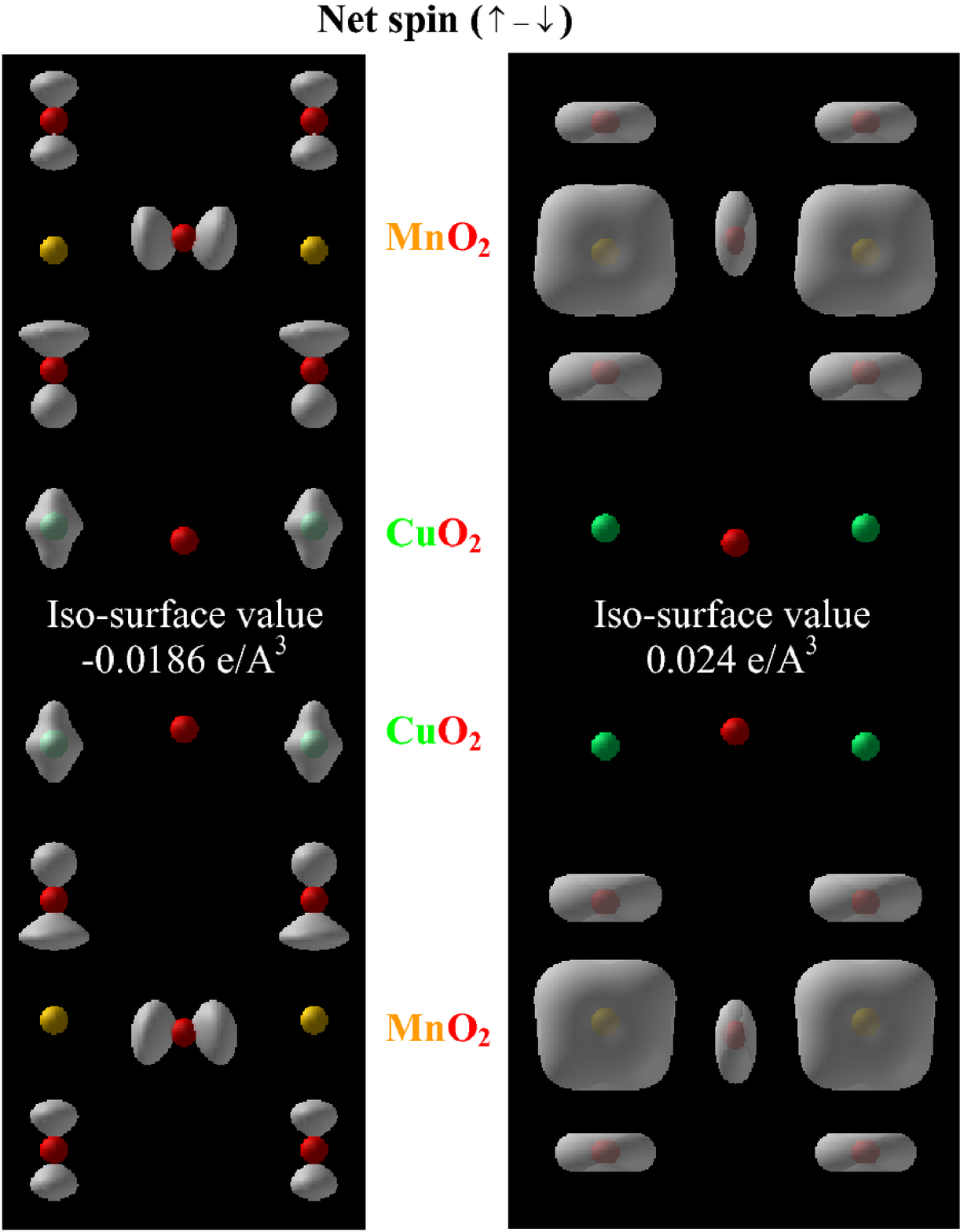}
\end{center}
\caption{\label{fig:SD}(Color online) The calculated spacial distribution of
  the spin-density in the $1\times$YBCO/$1\times$LMO super-lattice. Left and
  right panels show negative and positive constant-spin-density iso-surfaces,
  respectively. The cut is through a [100] plane, with $z$ axis parallel to
  $c$.}
 \end{figure}

The spacial distribution of the spin-density (SD) calculated for
the $1\times$YBCO/$1\times$LMO SL is presented in Fig.~\ref{fig:SD}. Large
positive iso-surfaces surrounding Mn ions (right panel of Fig.~\ref{fig:SD})
depict the spin-density of the completely spin-polarized Mn \tg\ states. The
Mn \dxyz--\OBa\ \pxy\ hybridization is responsible for the \textit{positive}
SD surface around \OBa. The left panel of Fig.~\ref{fig:SD} shows clearly
that \textit{negative} SD surfaces surrounding \OBa\ and Cu ions appear due
to the polarization of \OBa\ \pz\ and Cu \dzz\ states, respectively, which,
as it is explained above, is caused by their interaction with the spin-split
unoccupied Mn \dzz\ states.

Test calculations in which \OBa\ was shifted along the line connecting Mn and
Cu ions, while all other atomic positions were kept fixed, showed that the
calculated Cu moment is very sensitive to the position of the interfacial
\OBa\ ion. The variation of the Mn--\OBa\ and Cu--\OBa\ distances has a
twofold effect: first, the increase of the Mn--\OBa\ separation weakens the Mn
\dzz--\OBa\ \pz\ hybridization, which leads to a downward shift of the mostly
unoccupied \dzz\ bands relative to \dxxyy\ ones. The majority-spin \dzz\
states become more populated, which provides the dominant contribution to the
increase of the Mn spin moment from 2.75\mb, when the optimized \OBa\ position
is used ($d_{\text{Mn}-\text{O}_{\text{Ba}}}$=1.82 \AA), to 3.01\mb, when
\OBa\ is equidistant to the Mn and Cu sites
($d_{\text{Mn}-\text{O}_{\text{Ba}}}$=2.09 \AA). It should be noted, that in
spite of the increase of the Mn--\OBa\ distance, the negative
spin-polarization of the \pz\ states \textit{increases} as \OBa\ moves away
from Mn. This can be explained by the decrease of the energy separation
between the majority-spin Mn \dzz\ and \OBa\ \pz\ states as a result of the
downward shift of the former and an upward energy shift of the latter. Second,
strengthening of the Cu \dzz--\OBa\ \pz\ hybridization due to the decrease of
the Cu--\OBa\ distance increases the Cu \dzz\ weight in the Mn \dzz-derived
bands and, consequently, the spin-polarization of the Cu \dzz\ states. As
\OBa\ is shifted from the optimized position
($d_{\text{Cu}-\text{O}_{\text{Ba}}}$=2.37 \AA) to the middle of the Cu--Mn
line ($d_{\text{Cu}-\text{O}_{\text{Ba}}}$=2.09 \AA), \MCu\ increases from
$-$0.002 to $-$0.036\mb.

\subsection{\label{sec:cul23}Origin of XMCD at Cu $L_{2,3}$ edges}

\begin{figure}[tbp!]
\begin{center}
\includegraphics[width=0.45\textwidth]{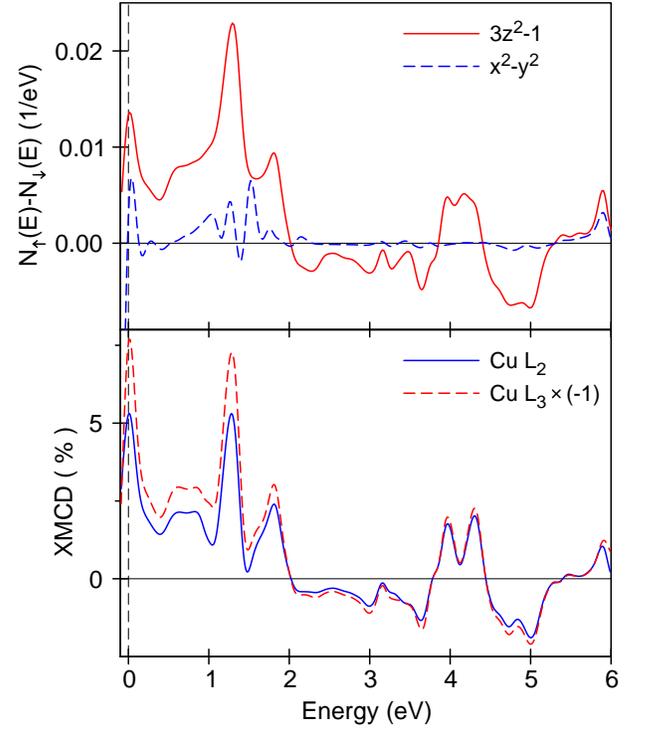}
\end{center}
\caption{\label{fig:l23z2} (Color online) The comparison of the calculated Cu
  $L_{2}$ and $L_{3}$ spectra (lower panel) and the difference of the
  densities of the majority- and minority-spin Cu \dzz\ and \dxxyy\ states
  (upper panel). The unbroadened spectra are plotted relative to the
  corresponding absorption edges.}
\end{figure}

The calculated Cu $L_{2}$ and $L_{3}$ XMCD spectra and the spin-polarization
of Cu \dzz\ and \dxxyy\ states, i.e., the difference of the densities of
corresponding majority- and minority-spin states, are plotted together in
lower and upper panels of Fig.~\ref{fig:l23z2}. In order to make the
similarities more clear, the spectra are plotted without the life-time and
instrumental broadening and are aligned to the corresponding absorption edges
which are shifted to zero energy. Figure \ref{fig:l23z2} shows unambiguously
that the energy dependence of the calculated XMCD signal at the Cu $L_{2}$ and
$L_{3}$ edges is determined by the spin-polarization of the Cu \dzz\ states.

\begin{table}[tbp!]
  \caption{\label{tab:sel_rule_L2} Squared angular matrix elements
    $\Omega^{2}_{\pm}$ for dipole allowed transitions at the $L_2$ 
    edge from an initial core state with the total angular momentum $j=1/2$
    and its projection $m_j$ to a final 3$d$ valence state ($|f\rangle$) with
    the absorption of left ($p=+1$) or right ($p=-1$) circularly polarized
    x-rays. $\Omega^{2}_{\pm}$ are multiplied by 90. 
    The intensity of unpolarized x-ray absorption and XMCD spectra are
    proportional to $\Omega^{2}_{+}+\Omega^{2}_{-}$ and
    $\Omega^{2}_{+}-\Omega^{2}_{-}$, respectively.}
\begin{ruledtabular}
\begin{tabular}{c/./...}
 & \multicolumn{2}{c}{$p=+1$} & \multicolumn{2}{c}{$p=-1$} \\
$|f\rangle$ & \multicolumn{1}{c}{$m_j$} & \multicolumn{1}{c}{$\Omega_{+}^2$} & 
\multicolumn{1}{c}{$m_j$} & \multicolumn{1}{c}{$\Omega_{-}^2$} & 
\multicolumn{1}{c}{$\Omega^{2}_{+}+\Omega^{2}_{-}$} &
\multicolumn{1}{c}{$\Omega^{2}_{+}-\Omega^{2}_{-}$} \\[0.5em] 
\hline
 \\
 \duyz     & -1/2 &  3 & -1/2 &  3 & 6 &  0 \\
 \duxz     & -1/2 &  3 & -1/2 &  3 & 6 &  0 \\
 \duxy     &  1/2 & 12 &      &    &12 & 12 \\
 \duzz     &      &    &  1/2 &  4 & 4 & -4 \\
 \duxxyy   &  1/2 & 12 &      &    &12 & 12 \\ [0.5em]
\hline 
\\
 \ddyz     &  1/2 &  3 &  1/2 &  3 & 6 &  0 \\
 \ddxz     &  1/2 &  3 &  1/2 &  3 & 6 &  0 \\
 \ddxy     &      &    & -1/2 & 12 &12 &-12 \\
 \ddzz     & -1/2 &  4 &      &    & 4 &  4 \\
 \ddxxyy   &      &    & -1/2 & 12 &12 &-12 \\
\end{tabular}
\end{ruledtabular}
\end{table}

\begin{table}[tbp!]
  \caption{\label{tab:sel_rule_L3} Squared angular matrix elements
    $\Omega^{2}_{\pm}$ for dipole allowed transitions at the $L_3$ 
    edge. The notations are the same as in Table \ref{tab:sel_rule_L2}.}  
\begin{ruledtabular}
  \begin{tabular}{c/./...}
 & \multicolumn{2}{c}{$p=+1$} & \multicolumn{2}{c}{$p=-1$} \\
$|f\rangle$ & \multicolumn{1}{c}{$m_j$} &
\multicolumn{1}{c}{$\Omega_{+}^2$} &  
\multicolumn{1}{c}{$m_j$} & \multicolumn{1}{c}{$\Omega_{-}^2$} & 
\multicolumn{1}{c}{$\Omega^{2}_{+}+\Omega^{2}_{-}$} & 
\multicolumn{1}{c}{$\Omega^{2}_{+}-\Omega^{2}_{-}$} \\ [0.5em]
\hline
\\
 \duyz   & -1/2 &  6 & -1/2 &  6 & 12 &   0 \\
 \duxz   & -1/2 &  6 & -1/2 &  6 & 12 &   0 \\
 \duxy   &  1/2 &  6 & -3/2 & 18 & 24 & -12 \\
 \duzz   & -3/2 &  6 &  1/2 &  2 &  8 &   4 \\
 \duxxyy &  1/2 &  6 & -3/2 & 18 & 24 & -12 \\ [0.5em]
\hline
\\
 \ddyz   &  1/2 &  6 &  1/2 &  6 & 12 &   0 \\
 \ddxz   &  1/2 &  6 &  1/2 &  6 & 12 &   0 \\
 \ddxy   &  3/2 & 18 & -1/2 &  6 & 24 &  12 \\
 \ddzz   & -1/2 &  2 &  3/2 &  6 &  8 &  -4 \\
 \ddxxyy &  3/2 & 18 & -1/2 &  6 & 24 &  12 \\
\end{tabular}
\end{ruledtabular}
\end{table}

Qualitative understanding of the Cu $L_{2,3}$ XMCD absorption spectra is
provided by the analysis of the matrix elements for $2p_{1/2,3/2} \to
3d_{3/2,5/2}$ transitions. XMCD at the $L_{2,3}$ edges is mostly determined by
the strength of the spin-orbit coupling of the initial 2$p$ core states and
spin-polarization of the final 3$d$ states while the exchange splitting of the
former as well as the SO coupling of the latter are usually of minor
importance. \cite{book:AHY04}
Squared angular matrix elements for the absorption of a photon with left
($p$=$+$1) or right ($p$=$-$1) circular polarization, $\Omega^{2}_{\pm}$, by
exciting an electron from the SO-split $p_{1/2,3/2}$ core levels to spin- and
crystal-field split $d$ states are collected in Tables \ref{tab:sel_rule_L2}
and \ref{tab:sel_rule_L3}. The matrix elements were calculated in the dipole
approximation neglecting the SO splitting of the $d$ states and assuming that
x-rays propagate along the spin quantization axis which is parallel to $z$.

By comparing Tables \ref{tab:sel_rule_L2} and \ref{tab:sel_rule_L3} we can
conclude that neglecting the radial matrix elements and different life-times
of 2$p_{1/2}$ and 2$p_{3/2}$ core holes, the unpolarized XAS intensity,
proportional to $\Omega^{2}_{+}+\Omega^{2}_{-}$, is twice as high at the $L_3$
edge as at the $L_2$ one. The corresponding XMCD signals,
$\Omega^{2}_{+}-\Omega^{2}_{-}$, are, however, of the same magnitude, with the
sign of the dichroism at the $L_2$ edge being opposite to the $L_3$
one. This can be clearly seen in the lower panel of Fig.~\ref{fig:l23z2} where
the unbroadened Cu $L_2$ and $L_3$ XMCD spectra are plotted. The almost
twofold difference in the intensities of the $L_2$ and $L_3$ XMCD spectra
presented in Fig.~\ref{fig:XMCD} appears because of the large difference in
the inverse life times of 2$p_{1/2}$ ($\Gamma_{L_2}$=1.04 eV) and 2$p_{3/2}$
($\Gamma_{L_3}$=0.61 eV) core holes. \cite{CaPa01}

The dependence of the matrix elements on the symmetry of the final $3d$ state
can be summarized as follows:
the dominant contributions to the x-ray absorption and to
the dichroism, i.e., the difference of the absorption probabilities for left
and right polarized photons, $\Omega^{2}_{+}-\Omega^{2}_{-}$, come from the
transitions to \dxy\ and \dxxyy\ states. The transitions to \dzz\ states are
weaker and their contribution to the dichroism is of the opposite
sign. Finally, the probabilities to excite a core 2$p$ electron to a \dxz\
or \dyz\ state with either left or right polarized photons are equal so that
these transitions do not contribute to the XMCD spectra.

It should be emphasized that the contributions to the dichroism from
transitions to final states with different spin projections have opposite
signs. Consequently, if majority- and minority-spin $d$ states with 
particular orbital character are degenerate their contributions to XMCD
spectra cancel each other. This is exactly what happens to the Cu \dxxyy\
contribution to the XMCD spectra calculated for the $1\times$YBCO/$1\times$LMO
SL: although the density of the unoccupied Cu \dxxyy\ states is much higher
than of the \dzz\ ones (see Fig.~\ref{fig:DOS_LSDA}) and, consequently,
transitions to the final states of the \dxxyy\ character control the x-ray
absorption at the $L_2$ and $L_3$ edges, the majority- and minority-spin
\dxxyy\ bands are degenerate and their contribution to the dichroism is very
weak.

In contrast to \dxxyy, the unoccupied Cu \dzz\ states, which, as explained in
Sec.~\ref{sec:bands}, follow the dispersion of the spin-split Mn \dzz\ bands,
are strongly polarized due to the Cu \dzz--\OBa\ \pz--Mn \dzz\ hybridization.
As the minority-spin Mn bands are shifted to the energies above 2 eV by the
strong exchange interaction within the Mn $d$ shell (see
Fig.~\ref{fig:BND_FB}), only the majority spin Cu \duzz\ holes are created in
the energy range up to 2 eV above \ef. Although the number of the \duzz\
holes is very small, it is sufficient to produce the appreciable XMCD signal
at Cu $L_{2,3}$ edges.

According to Tables \ref{tab:sel_rule_L2} and \ref{tab:sel_rule_L3} the matrix
elements for the transitions to the final states of the \dzz\ character are
small. It should be noted, however, that they were derived for the light
propagation and magnetization directions parallel to $z$, i.e., to the
tetragonal $c$ axis, whereas the XMCD spectra shown in Fig.~\ref{fig:XMCD}
were calculated for the experimental geometry \cite{CFS+06} in which
circularly polarized x-rays propagate along the direction of the Mn
magnetization which is parallel to the $a$ axis. The dependence of the XMCD
spectra on the light propagation direction can be estimated by expanding the
Cu \dzz\ orbital in terms of cubic harmonics defined in the rotated frame with
$z'||a$:
\[
d_{z^2} = -\frac{1}{2}d'_{z^2}+\frac{\sqrt{3}}{2}d'_{x^2-y^2} \,.
\]
Because of the appearance of the $d'_{x^2-y^2}$ contribution in the expansion
of the corresponding wave functions, the Cu $L_{2,3}$ XMCD spectra calculated
in the $z'||a$ geometry should change sign and become more intense compared
to the $z||c$ spectra. This strong orientational dependence of the Cu XMCD
spectra was verified by performing calculations for the
$1\times$YBCO/$1\times$LMO SL with $\MMn||c$. Indeed, due to the weakness of
the matrix elements for the transitions to the \dzz-like final states, the
magnitude of the Cu $L_{2,3}$ XMCD spectra calculated in this geometry is
smaller and their signs are opposite compared to the spectra shown in
Fig.~\ref{fig:XMCD} which were obtained with the light propagation and \MMn\
directions parallel to $a$.

\subsection{\label{sec:robust}Robustness of calculated Cu $L_{2,3}$ XMCD}

The proposed explanation for the strong XMCD at the Cu $L_{2,3}$ edges and
robustness of the calculated spectra have been confirmed by a number of
additional calculations.
As explained in Sec.~\ref{sec:bands}, the Cu magnetization induced by the
proximity to the LMO layer is sensitive to the position of \OBa. This is
also true for the XMCD spectra. The increase of the Cu \dzz--\OBa\ \pz\
hybridization strength with the decrease of the Cu--\OBa\ distance leads to
the growth of the Cu \dzz\ weight in the Mn \duzz-derived bands. Because of
the increased number of Cu \duzz\ holes the calculated Cu $L_{2,3}$ XMCD
spectra become more intense as \OBa\ shifts from Mn toward Cu.

As was mentioned above the number of electrons occupying Cu ($d^{9\pm\delta}$)
and Mn ($d^{3\mp\delta}$) 3$d$ shells in the model $1\times$YBCO/$1\times$LMO
SL corresponds to the composition of the La$_{1-x}$Ca$_x$MnO$_3$ layer with
$x\sim 1$. By using the virtual crystal approximation we performed also LSDA
calculations with the increased number of valence electrons corresponding to
Ca doping of $x$=2/3 and 1/3. The added electrons fill mainly Cu \dxxyy\
and majority-spin Mn \dxxyy\ bands crossing the Fermi level. Due to the
increase of the Fermi energy, the Mn \duzz-derived bands moves closer to \ef\
but their occupation does not change. As a consequence, the change of the
electron count has only a very weak effect on the Cu XMCD spectra.

Finally, the effect of the increase of the LMO layer thickness on the Cu
$L_{2,3}$ XMCD spectra was studied by performing LSDA calculations for
$1\times$YBCO/$n\times$LMO ($n$=2, 3) superlattices with FM alignment of Mn
magnetic moments. A LMO layer in these SL consists of 2 and 3 unit cells of
pseudo-cubic LaMnO$_3$, respectively, i.e., $3\times$MnO$_2$+$2\times$LaO and
$4\times$MnO$_2$+$3\times$LaO planes. The distance between the MnO$_2$ planes
was fixed to $c_{\mathrm{LMO}}$=3.94 \AA. For ions at the interface the
positions obtained from the structure optimization for the
$1\times$YBCO/$1\times$LMO SL were used. In contrast to the latter, in which
both MnO$_2$ planes are at the interface with the YBCO layer, the thicker
superlattices have additionally one ($n$=2) or two ($n$=3) ``bulk'' MnO$_2$
plane(s). Mn ions in these bulk MnO$_2$ planes are surrounded by an
undistorted $O_6$ octahedron, with all the Mn--O distances of 1.97 \AA\ being
equal to that in pseudo-cubic LaMnO$_3$.

Assuming that the Cu valency does not depend on the LMO layer thickness, the
effective Mn $d$ electron count in these SL increases compared to the SL with
$n$=1 and corresponds to Ca content of $x$=2/3 for $n$=2 and $x$=1/2 for
$n$=3. In order to separate the effects of the LMO layer thickness and that of
the varying occupation of the Mn 3$d$ shell, XMCD calculations for all three
SL were performed for the same effective Mn $d$ occupation corresponding to
$x$=2/3.

The magnetic moment of $\sim$3.2\mb\ calculated for the bulk Mn ions is higher
than for the interfacial ones and close to the LSDA value obtained for
pseudo-cubic La$_{1-x}$Ca$_x$MnO$_3$ with the same $x$=2/3. The moment of the
interfacial Mn ions and the magnetization induced on Cu remain nearly constant
independently of $n$. As a consequence, the Cu $L_{2,3}$ XMCD spectra
calculated for the SL with $n$=1, 2, and 3 are nearly identical, which shows
that they are not sensitive to the LMO layer thickness.

\subsection{\label{sec:ldau}Effect of electronic correlations}

The influence of electronic correlations on the Cu $L_{2,3}$ XMCD spectra
calculated for the $1\times$YBCO/$1\times$LMO SL was investigated using the
rotationally invariant LSDA+$U$ method.\cite{LAZ95,YAF03} In order to
understand the LSDA+$U$ results it is worth recalling that the expression for
the additional orbital dependent LSDA+$U$ potential $V_i$ acting on $i$-th
localized orbital becomes particularly simple if non-spherical contributions
to the screened Coulomb $U$ and exchange $J$ integrals are neglected:
\[
V_i=(U-J)(\frac{1}{2}-n_i) ,
\]
where $n_i$ is the occupation of the $i$-th orbital. Thus, the occupied
(unoccupied) states are shifted downwards (upwards) by $\sim(U-J)/2$.

When $U$ is applied only to the Mn $d$ states its main effect is to increase
the splitting between the occupied majority- ($n_{i}\sim 0.9$) and unoccupied
minority-spin ($n_{i}\sim 0.1$) Mn ``\tg'' states. This causes some increase
of \MMn\ from the LSDA value of 2.8\mb\ to 3.0\mb\ for $U$=3 eV and 3.3\mb\
for $U$=5 eV. The minority-spin \dzz\ and \dxxyy\ states with $n_{i}\sim 0.2$
are also shifted to higher energies.
The occupation numbers of the majority-spin Mn \dzz\ and \dxxyy\ states are
close to 0.5 because of their strong hybridization with O $p$ states and
partial filling of the corresponding bands (see Fig.~\ref{fig:BND_FB}). As a
consequence the majority-spin Mn \dzz- and \dxxyy-derived bands are almost
unaffected by $U$. Hence, the Cu $L_{2,3}$ XMCD spectra, which are mostly
determined by the Cu \duzz -- Mn \duzz\ hybridization, show weak dependence on
$U$ applied to the Mn $d$ states.

LSDA+$U$ calculations with $U$=7.5 eV acting within the Cu 3$d$ shell were
performed for an $\sqrt{2}a\times \sqrt{2}a\times c$ unit cell of the
$1\times$YBCO/$1\times$LMO SL. In this enlarged cell two Cu sites in each
CuO$_2$ plane become inequivalent which enables calculations with
antiferromagnetic (AFM) ordering of Cu moments. The calculations were
performed for a collinear magnetic arrangement with both Cu and Mn magnetic
moments parallel to the tetragonal $a$ axis.

LSDA calculations for the double cell converge to the same magnetic solution
which was obtained for the single cell and is characterized by FM ordering of
Mn moments and a tiny FM moment induced across the interface on Cu
ions. However, when $U$=7.5 eV is applied to the Cu $d$ states and AFM
ordering within a CuO$_2$ plane is allowed, Cu ions acquire a spin moment of
$\sim$0.4\mb. The moment appears due to splitting of the partially occupied
majority- and minority-spin Cu \dxxyy\ states by the on-site Coulomb
repulsion. The other Cu $d$ states, including \dzz\ ones, which are completely
filled already in LSDA calculations, are shifted $\sim$3 eV downwards to the
bottom of O $p$ bands.

Because of the proximity to FM MnO$_2$ planes the Cu moments which are
parallel (0.44\mb) and antiparallel ($-$0.45\mb) to \MMn\ are slightly
different, with the difference being comparable to the LSDA moment induced in
the Cu \dzz\ chanel by the Cu \dzz--\OBa\ \pz--Mn \dzz\ hybridization as
discussed in Sec.~\ref{sec:bands}.
Although the hybridization causes a small ($\sim$0.01\mb) difference in the
moments of two inequivalent Mn ions which have Cu neighbors with either
parallel or antiparallel magnetization, the calculated \MMn\ of 2.75\mb\ and
2.76\mb\ are very close to the LSDA value.

\begin{figure}[tbp!]
\begin{center}
\includegraphics[width=0.45\textwidth]{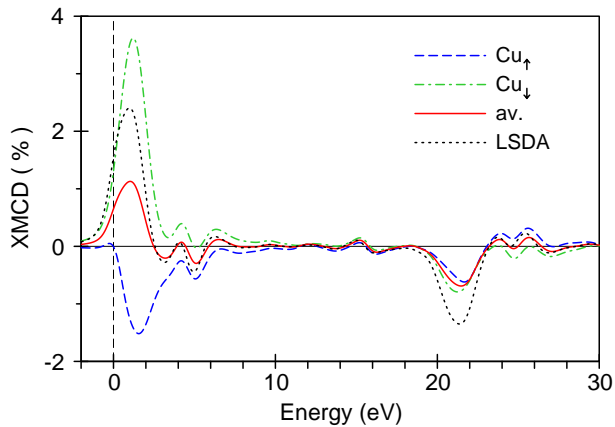}
\end{center}
\caption{\label{fig:xmcd_Cu_af} (Color online) Cu $L_{2,3}$ XMCD spectra
  calculated using the LSDA+$U$ method with $U$=7.5 eV applied to the Cu 3$d$
  states for Cu ions with the magnetic moment parallel (dashed line) and
  antiparallel (dash-dotted line) to \MMn. The averaged spectrum is plotted by
  solid line. For comparison, the LSDA spectrum from Fig.~\ref{fig:XMCD} is
  also shown (dotted line). The spectra are plotted relative to the $L_3$
  absorption edge.}
\end{figure}

Figure~\ref{fig:xmcd_Cu_af} shows that the spin-polarization of the Cu \dxxyy\
states affects strongly the $L_{2,3}$ XMCD spectra calculated for Cu ions with
the moment parallel (Cu$_{\uparrow}$) and antiparallel (Cu$_{\downarrow}$) to
\MMn. As the majority- and minority-spin Cu \dxxyy\ states are split by the
on-site Coulomb repulsion, transitions to the \dxxyy\ final states give a
substantial contribution to the dichroism. In the geometry with the x-ray
propagation and magnetization directions parallel to the $c$ axis the
contribution of the \dxxyy\ states would be the dominant one and the
Cu$_{\uparrow}$ and Cu$_{\downarrow}$ XMCD spectra would be of the same shape
but opposite signs. Since the calculations were performed with $\MCu||a$ the
matrix elements for the transitions to the \dxxyy\ states are strongly reduced
and their contribution is comparable to the \dzz\ one, which is induced by the
hybridization with FM ordered Mn and, thus, is of the same sign for both Cu
ions. As a result, the shapes of the Cu$_{\uparrow}$ and Cu$_{\downarrow}$
$L_{3}$ XMCD spectra are different (Fig.~\ref{fig:xmcd_Cu_af}), whereas the
$L_2$ spectra are of the same sign although the Cu$_{\uparrow}$ and
Cu$_{\downarrow}$ moments are antiparallel. Nevertheless, when the individual
Cu$_{\uparrow}$ and Cu$_{\downarrow}$ spectra are averaged, the net Cu XMCD
spectrum becomes similar to the LSDA spectrum calculated for the single
$1\times$YBCO/$1\times$LMO cell, which is also shown in
Fig.~\ref{fig:xmcd_Cu_af} for comparison sake. The magnitude of the averaged
spectrum is two times smaller, which may be explained by decrease in the
weight of the Cu \dzz\ states in the minority-spin Mn \dzz\ bands caused by
their downward shift in the LSDA+$U$ calculations. Hence, even in the presence
of the sizable Cu moment the dichroic signal at the Cu $L_{2,3}$ edges is
still governed by the spin-polarization of the Cu \dzz\ states induced by the
Cu \dzz--Mn \dzz hybridization via the $p$ states of interfacial \OBa\ ions.

Note, that if the AFM aligned Cu \dxxyy\ moments were oriented perpendicular
to the Mn magnetization and, consequently, to the x-ray propagation direction
they would not contribute to the dichroism, which also in this case would be
determined by the transitions to the Cu \dzz\ states.

\section{\label{sec:summ}Conclusions}

We studied the electronic structure, XAS and XMCD spectra of YBCO/LMO
superlattices by means of the {\it ab initio} PAW and fully-relativistic
spin-polarized LMTO methods.

Our results contradict to the orbital reconstruction model, proposed in
Ref.~\onlinecite{CFH+07} in order to explain weak XLD at the Cu $L_{2,3}$
edges measured in the TEY mode. In all the calculations performed in the
present work the number of Cu \dzz\ holes is significantly smaller than the
number of \dxxyy\ ones. The large difference in the number of \dzz\ and
\dxxyy\ holes causes strong linear dichroism of the calculated Cu $L_{2,3}$
XAS.

On the other hand, the LSDA calculations reproduce very well the experimental
Cu $L_{2,3}$ XMCD spectra of Ref.~\onlinecite{CFS+06}. We have shown that the
theoretical XMCD spectra are proportional to the difference of the densities
of unoccupied majority- and minority-spin Cu \dzz\ states. Although the Cu
\dzz-derived bands lie $\sim$2 eV below the Fermi level, a small number of
\dzz\ holes is created because of the Cu \dzz--\OBa\ $p_z$--Mn \dzz\
hybridization, which induces a small Cu magnetic moment opposite to the moment
of Mn ions. However, this tiny number of \dzz\ holes is sufficient to produce
substantial XMCD at the Cu $L_{2,3}$ edges.

The calculated Cu $L_{2,3}$ XMCD spectra are very robust: they are not
sensitive to Ca doping of the La$_{1-x}$Ca$_x$MnO$_3$ layer simulated using
the virtual crystal approximation, nor to the LCMO layer
thickness. Remarkably, the shape of the XMCD spectra obtained from LSDA+$U$
calculations does not change much even when $U$=7.5 eV is applied to the Cu
$d$ states and AFM order with the Cu moment of 0.4\mb\ is induced in the
CuO$_2$ plane.

\section*{Acknowledgments}

Authors are thankful to G.~Khaliullin, C.~Bernhard, H.-U.~Habermeier, and
J.~Chakhalian for helpful discussions.
V.N. Antonov gratefully acknowledges the hospitality at Max-Planck-Institut
f\"ur Festk\"orperforschung in Stuttgart during his stay there. 
This work was partially supported by Science and Technology Center in Ukraine
(STCU), Project No. 4930.

\newcommand{\noopsort}[1]{} \newcommand{\printfirst}[2]{#1}
  \newcommand{\singleletter}[1]{#1} \newcommand{\switchargs}[2]{#2#1}

\end{document}